\def   \ni {\noindent}

\def   \ssk {\vskip  5truept}

\def   \bsk {\vskip 15truept}

\def   \newline {\hfil\break}

\documentstyle[epsfig]{article}
\begin{document}

\hsize 5truein
\vsize 8truein
\font\abstract=cmr8
\font\keywords=cmr8
\font\caption=cmr8
\font\references=cmr8
\font\text=cmr10
\font\affiliation=cmssi10
\font\author=cmss10
\font\mc=cmss8
\font\title=cmssbx10 scaled\magstep2
\font\alcit=cmti7 scaled\magstephalf
\font\alcin=cmr6 
\font\ita=cmti8
\font\mma=cmr8
\def\ref{\par\noindent\hangindent 15pt}
\null
\pagestyle{empty}


\title{\ni OBSERVATIONS OF THE GALACTIC CENTER \\
REGION WITH BEPPO-SAX}                                               

\bsk \bsk
\author{\ni L.~Sidoli $^{1,2}$, S.~Mereghetti$^{2}$, L.~Chiappetti$^{2}$, J.~Heise$^{3}$, 
G.L.~Israel $^{4}$, E.~Kuulkers$^{3}$,  M.~Orlandini$^{5}$,
P.~Predehl$^{6}$, A.~Tiengo$^{1,2}$ \&   A.~Treves$^{7}$ 
}                                                       
\bsk
\affiliation{1) Dipartimento di Fisica, Universit\`a di Milano, Milano, Italy   \\
\indent {2) Istituto di Fisica Cosmica ``G. Occhialini" , Milano, Italy} \\
\indent {3) SRON-Utrecht, Utrecht, The Netherlands} \\
\indent {4) Osservatorio Astronomico di Roma, Monteporzio Catone, Roma, Italy} \\
\indent {5) TeSRE, Bologna, Italy} \\
\indent {6) MPE, Garching, Germany} \\
\indent {7) Dipartimento di Fisica, Universit\`a di Milano, sede di Como, Italy} \\
  
}                                                
\bsk
\baselineskip = 12pt

\abstract{ABSTRACT \ni

We are performing  a survey of the Galactic Center 
region (Sidoli et al., 1998a) with the BeppoSAX 
satellite.  Several known point sources are 
visible (including one at 
the position of SgrA*),
as well as newly discovered sources and  diffuse emission.
Here we report the  preliminary results of the on--going 
analysis of both the
point sources   and   the  diffuse X--ray emission. 

}                                                    
\bsk
\baselineskip = 12pt

\bsk
\baselineskip = 12pt


\text{\ni 1. INTRODUCTION
\ssk
\ni     

The region ($ |l| < 2^\circ $})$\times$($|b| < 2^\circ $) around the 
Galactic Center was observed with the BeppoSAX satellite (Boella et al., 1997) 
during April 1997--1998 and August--September 1997 
for a total of $\sim$120 hours of
effective time exposure. An observation of the black hole
candidate GRS1758--258 has also been obtained on April 10, 1997. 
Figure 1 shows a mosaic of the MECS images obtained 
during this project.   
In Table 1 we report the preliminary results of 
the spectral analysis of the point sources observed by the MECS instruments 
in the energy band 2--10 keV.  
Details on the analysis performed will be reported elsewhere 
(Sidoli et al., in preparation).

\baselineskip = 12pt

\bsk
\ni 2. INDIVIDUAL SOURCES 
\ssk
\ni  

X--ray emission from the core of the radio supernova remnant 
G0.9+0.1 was discovered 
in April 1997 (Mereghetti et al., 1998), confirming the 
composite morphology derived from the radio observations.
A young, energetic pulsar in G0.9+0.1 could contribute to the 
unresolved gamma--ray excess observed by EGRET 
in the Galactic Center (Mayer-Hasselwander et al., 1998).

 
\begin{figure}[h]
\vskip .5truecm
\centerline{\psfig{file=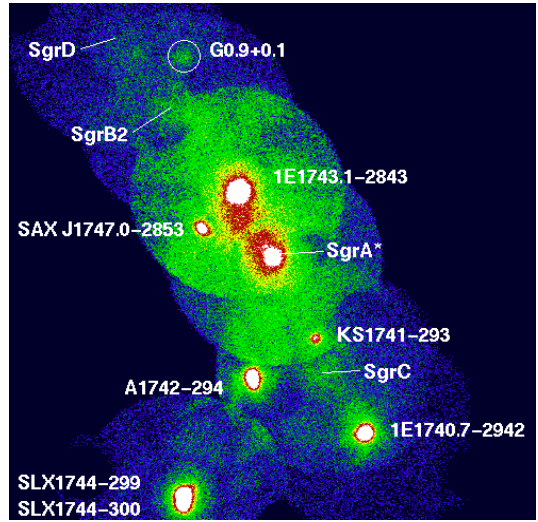, width=7cm}}
\vskip 0.truecm
\caption{FIGURE 1. Image of the Galactic Center region 
($ |l| < 2^\circ $) in the 2--10 keV energy range, not corrected for 
the exposure. North is to the top and East to the left.}
\end{figure}



\begin{center}
\vskip 2.truecm
\small
{\centerline{  TABLE 1. INDIVIDUAL SOURCES: MECS Results of the 
Spectral Analysis (errors are 90\% c.l.).}}
\begin{tabular}[c]{cccccc} 
 \hline
Source    & Class$^{a}$   & Obs. date   & $N_{H}$      & Parameter$^{b}$   & Flux$^{c}$      \\
         &       	  &          	& ($10^{22}$ cm$^{-2}$) &     of best fit         &    \\
\hline 
G0.9+0.1 & SNR		& Apr/Sep 97	& $34^{+16}_{-9}$   		& $\Gamma=3.7^{+1.3}_{-1.0}$     & $1.36^{+3.74}_{-0.72}$  \\  
Galactic Center$^{d}$	&  UN& Aug 1997	& $7.0\pm{0.3}$  	 & $kT_{M}= 4.1\pm{0.3} $  & $3.5\pm{0.2}$  \\
1E1743.1--2843 &  UN  &	  Apr 1998 	& $13\pm{0.5}$   	  	& $kT_{bb}= 1.78\pm{0.03} $  & $16.5\pm0.3$                 \\
SAX~J1747.0--2853 & NS  & Apr 1998	& $8.3\pm{0.6}$   	& $kT_{br}=6.1\pm{0.9} $            & $4.0\pm{0.3}$      \\ 
KS1741--293 & NS	& Mar 1998	& $20\pm{0.2} $  	  	& $kT_{br}=11\pm{3}$            & $13\pm{1}$            \\  
1E1740.7--2942 & BHC & 	Sep 1997	& $14.7\pm{0.4}$    	& $\Gamma=1.52\pm{0.04}$  &  $47.4\pm{0.04}$   \\
SLX1744--299 & NS &  	Sep 1997	&$5.1\pm{0.2}$            	 & $\Gamma=2.1\pm{0.1}$    & $20\pm{0.4}$   \\
SLX1744--300 & NS &  	Sep 1997	&$5.3\pm{0.2}$            	 & $\Gamma=2.2\pm{0.1}$    &  $12\pm{0.3}$               \\
GRS1758--258 & BHC & 	Apr 1997	&$1.44\pm{0.1}$     	  & $\Gamma=1.55\pm{0.03}$ & $30\pm{0.2}$               \\

\end{tabular}
\end{center}
\begin{small}
$^{a}${Class: BHC=black hole candidate; NS=neutron star; SNR=supernova remnant; UN=unknown};
$^{b}${ Power law photon index ($\Gamma$) or temperatures in keV for a  black body (T$_{bb}$), 
bremsstrahlung (T$_{br}$) and emission  from hot gas (T$_{M}$, MEKAL in XSPEC).} \\
$^{c}${Unabsorbed fluxes are in the energy band 2--10 keV in units of $10^{-11}$ ergs~cm$^{-2}$~s$^{-1}$}.\\
$^{d}${The total contribution of the sources  within $\sim2'$ from the Galactic Center (see text).}\\
\end{small}


\vskip 2.truecm 
 
\indent
A type I X--ray burst from SAX~J1747.0--2853, an X--ray 
transient recently 
rediscovered  with the WFC on-board BeppoSAX (in't Zand et 
al. 1998b, Bazzano et al. 1998) and positionally
coincident with the X--ray transient GX~0.2--0.2,  has been detected 
with both MECS (Sidoli et al., 1998b)
and PDS  on April 15, 1998. The burst light curve 
is shown in Fig. 2   as well as the luminosity, 
temperature and neutron star radius variations during the burst.

 \begin{figure}[h]
\vspace{2.9truecm}
\psfig{figure=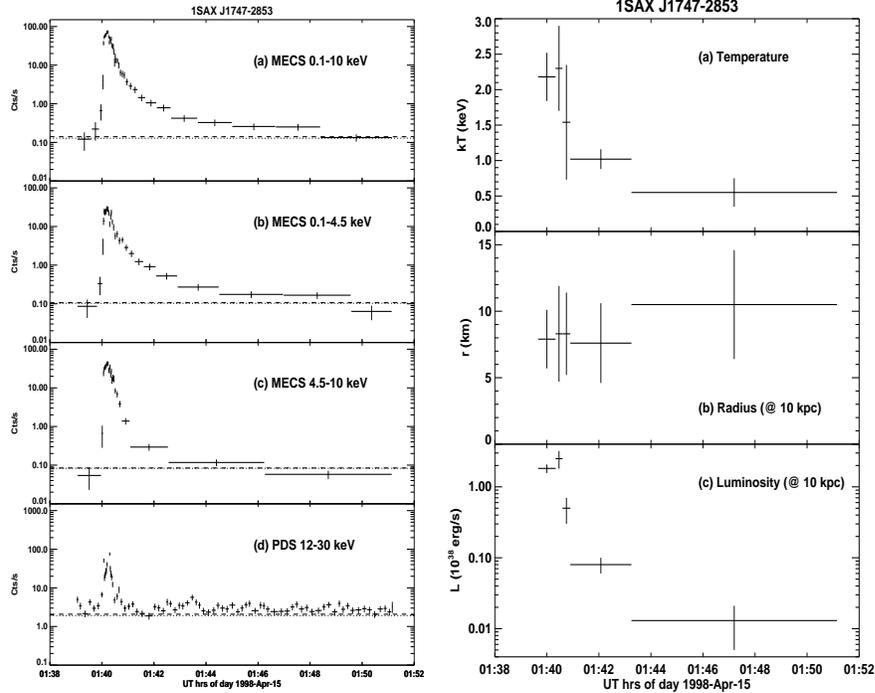,height=120mm,bbllx=15mm,bblly=60mm,bburx=140mm,bbury=220mm}

\vskip -5.6truecm

\caption{\small FIGURE 2. {\underline{Left panel}}: Light curve of the burst detected  
with the MECS and PDS from SAX~J1747.0--2853. The dotted and dashed 
lines  indicate the level of the persistent 
emission in the contiguous interval respectively before and after 
the burst. {\underline{Right panel}}: Results obtained by 
fitting the burst emission in different time intervals with a blackbody 
spectrum.    
}

\end{figure}

\vskip -0.3truecm

\indent
The transient activity of a source positionally coincident with 
the burster KS1741--293 (in't Zand et al., 1998a) has been 
detected during the observation 
pointed on the molecular cloud SgrC. In March 1998 it had an 
X-ray luminosity (2--10 keV) L$_{X}\sim10^{36}$ erg~s$^{-1}$, while 
in September 1997 it was under the threshold of detectability 
(L$_{X}$ $<$ 10$^{35}$ erg~s$^{-1}$).The 0.1--200 keV spectra 
of the two black hole candidates 1E1740.7--2942 and 
GRS1758--258 have been fitted using LECS, MECS and PDS data. 1E1740.7--2942, observed 
in September 
1996 and exactly one year later, did not show strong flux variability. 
For both sources a Sunyaev and Titarchuk comptonization
model fitted the data better than a simple power law. For 1E1740.7--2942 the  
parameters are: N$_{H}=1.46\pm{0.02}\times10^{23}$ cm$^{-2}$, 
$T = 24\pm1$ keV, $\tau=5.5\pm0.1$  and
unabsorbed flux F$_{2-10}\sim5\times10^{-10}$ ergs~cm$^{-2}$~s$^{-1}$
($\chi^{2}=1.3$, 282 d.o.f.). The 
corresponding values 
for GRS1758--2588 are: N$_{H}=0.18\pm{0.01}\times10^{23}$ cm$^{-2}$, 
$T =32\pm{4}$ keV, $\tau=4\pm{0.3}$ and 
F$_{2-10}\sim3.5\times10^{-10}$ ergs~cm$^{-2}$~s$^{-1}$ ($\chi^{2}=1.05$, 748 d.o.f.).

\baselineskip = 12pt

\bsk
\ni 3. SgrA. 
\ssk
\ni  
The possible existence of an X--ray counterpart of the radio source 
SgrA*, the dynamical center of the Galaxy, is still an open issue.
The Einstein Observatory   source 1E~1742.5--2859 (Watson et al., 1981) was 
later resolved
into three sources by ROSAT PSPC (Predehl \& Trumper, 1994). One of these 
sources, highly absorbed and located within
$10''$ from Sgr A*, is very likely associated with it. About 
$1.5'$ SW of SgrA* ASCA revealed
a harder source, possibly the quiescent state
of the bright soft X-ray transient A1742-289 (Maeda et al. 1996, but see
also  Kennea \& Skinner 1996).
The flux reported in Table 1 refers to the total contribution of 
the sources  within $\sim2'$ from the Galactic Center.  Indeed the 
BeppoSAX spatial resolution hampers a detailed analysis of 
the single sources present in this region.

\bsk
\ni 4. DIFFUSE X--RAY EMISSION.
\ssk
\ni
The  MECS spectrum of the diffuse 
emission  in a circular corona from $2'$ to  $8'$ around the Galactic 
Center is thermal (kT$\sim$7~keV, 
N$_{H}\sim4\times10^{22}$ cm$^{-2}$) 
and contains several emission lines, with the K-lines 
from iron and sulfur particularly bright. The corresponding luminosity 
is $\sim$$10^{36}$ ergs  s$^{-1}$, while
the luminosity of the other three molecular clouds (SgrB2, SgrC and SgrD)
within $8'$ from the center of their emission ranges from 0.1 
to 0.3$\times$$10^{36}$ ergs  s$^{-1}$. The  X--ray diffuse emission 
detected from the direction of these 
molecular clouds is harder and the intensity of the emission lines 
weaker with respect to  SgrA.
A 6.4 keV fluorescent iron line is detected and is found to be 
particularly strong in the direction of the molecular cloud SgrB2.

\bsk
\baselineskip = 12pt

{\references \ni REFERENCES
\ssk

\ref A. Bazzano et al. 1998, IAU Circ. n.6873
\ref G. Boella, R.C. Butler, G.C. Perola et al., 1997, A\&AS 122, 299
\ref J.A. Kennea \& G.K. Skinner, 1996, PASJ 48, 117
\ref Y. Maeda et al., 1996, PASJ 48, 417
\ref H.A. Mayer-Hasselwander et al., 1998, A\&A 335, 161  
\ref S. Mereghetti , L. Sidoli \& G.L. Israel, 1998, A\&A, 331, L77  
\ref P. Predehl \& J. Trumper 1994, A\&A 290, L29
\ref L. Sidoli et al., 1998a, proc. XTE/SAX Symp. ``The Active X-ray Sky", Rome , 88
\ref L. Sidoli et al., 1998b, A\&A, 336, L81
\ref M.G. Watson et al., 1981, ApJ 250,142 
\ref J.J. in't Zand et al., 1998a, IAU circ. n.6840  
\ref J.J. in't Zand et al. 1998b, IAU circ. n.6846

}                      

\end{document}